\documentclass[12pt]{article}
\usepackage{amsmath}
\usepackage{citesort}
\usepackage{amssymb}
\usepackage{epsfig}
\def \BE {\begin{equation}}
\def \EE {\end{equation}}
\def \BEA {\begin{eqnarray}}
\def \EEA {\end{eqnarray}}

\begin{document}

\title{Kelvin-wave turbulence generated by vortex reconnections.}
\author{Sergey Nazarenko
\\ \ \\
{Mathematics Institute,  University of Warwick, Coventry CV4 7AL, UK}}
\maketitle

\begin{abstract}
Reconnections of quantum vortex filaments create sharp bends
which degenerate into propagating Kelvin waves.
These waves cascade their energy down-scale and 
their waveaction up-scale via weakly nonlinear interactions, 
and this is the main mechanism of turbulence at the scales
less than the inter-vortex distance. 
In case of an idealised  forcing concentrated 
around a single scale $k_0$, the turbulence spectrum exponent
has a
pure direct cascade form $-17/5$ at scales $k>k_0$
\cite{svist} and  a pure inverse cascade form
$-3$ at $k<k_0$ \cite{leb}. However, forcing produced by the 
reconnections contains a broad range of Fourier modes.
What scaling should one expect in this case?
In this Letter I obtain an answer to this question using
the differential model for the  Kelvin wave turbulence introduced
in \cite{naz_kelvin}.
The main result is that the direct cascade scaling dominates,
i.e. the reconnection forcing is more or less equivalent
to a low-frequency forcing.

\end{abstract}

\section{Differential equation model for Kelvin wave turbulence.}

Superfluid turbulence, when excited at scales much greater
than the mean separation between quantum vortices, behaves
similarly to turbulence in classical fluids at such
large scales in that it develops a Richardson-like cascade
characterised by Kolmogorov spectrum \cite{vinen_intro}.
However,
quantum turbulence starts feeling discreteness when the energy
cascade reaches down to the length-scales comparable to the mean
inter-vortex separation distance.
In superfluids
near zero temperature, there is no normal component and, therefore, there
is no viscid of frictional dissipation in the system.
Even though part of the turbulent energy is lost to sound
radiation during the vortex reconnection processes, the 
major part of it is believed to be continuing to cascade to the
scales below the inter-vortex separation scale via nonlinear
interactions of Kelvin waves
\cite{svist,vinen,kivot,ks,vtm,ks1}.
Following  \cite{naz_kelvin}
I will refer to this state characterised by random nonlinearly interacting
Kelvin waves as ``kelvulence'' (i.e. Kelvin turbulence).
Kozik and Svistunov \cite{ks} used the weak turbulence
approach to kelvulence and derived a six-wave kinetic equation (KE) for the
spectrum of weakly nonlinear Kelvin waves. Based on KE, they derived a
spectrum of waveaction that corresponds to the constant Kolmogorov-like
cascade of energy from small to large wavenumbers,
\begin{equation}
n_k \sim k^{-17/5}.
\label{direct}
\end{equation}
Because the number of waves in the leading resonant process is even (i.e. 6),
KE conserves not only the total energy but also the total waveaction of the system.
The systems with two positive conserved quantities are known in turbulence 
to possess a dual cascade behaviour. For the Kelvin waves, besides the direct
energy cascade there also exists an inverse cascade of waveaction \cite{leb},
\begin{equation}
n_k \sim k^{-3}.
\label{inverse}
\end{equation}
Numerical confirmation of the direct cascade spectrum (\ref{direct})
was given by
Kozik and Svistunov \cite{ks1} who forced the system at the largest
scales. To date, there has been no simulations with forcing concentrated
at the smallest scales and, therefore, there is no numerical confirmation
of the inverse cascade spectrum.

On the other hand, in superfluids kelvulence is generated by vortex reconnections
which is not concentrated in either large or small scales, but it has a
continuous $k$-space distribution. Indeed, a sharp bend on the vortex line
produced by a reconnection has spectrum $n_k \sim k^{-4}$. What scaling should
we expect in kelvulence pumped by the reconnections, - forward cascade, inverse
cascade or a mixture of thereof? In the present paper I will answer this question
using a differential approximation model (DAM) introduced in
\cite{naz_kelvin},
\begin{equation}
\dot n = {C  \over \kappa^{10}} \omega^{1/2} {\partial^2 \over \partial \omega^2}
\left(
 n^6 \omega^{21/2} {\partial^2 \over \partial \omega^2} {1 \over n}
\right) + F_k -D_k,
\label{forth}
\end{equation}
where $\kappa$ is the vortex line circulation, $C$ is a dimensionless constant
and $\omega = \omega (k) = {\kappa \over 4 \pi} k^2$ is the Kelvin wave
frequency (we ignore logarithmic factors). Here $F_k$ and $D_k$ are the
terms describing forcing and dissipation of Kelvin waves.

In absence of forcing and dissipation, $F_k=D_k=0$, DAM  preserves the energy
\begin{equation}
E = \int \omega^{1/2} n \, d\omega
\end{equation}
and the waveaction
\begin{equation}
N = \int \omega^{-1/2} n \, d\omega.
\end{equation}
In this case equation (\ref{forth}) has both the
direct cascade solution (\ref{direct}) and  the inverse cascade solution
 (\ref{inverse}).
It also has thermodynamic Rayleigh-Jeans solutions,
\begin{equation}
n = { T \over  \omega + \mu}.
\label{term}
\end{equation}
where $T$ and $\mu$ are constants having a meaning of temperature and the chemical
potential respectively. 

Now let us assume that the forcing is due to the vortex reconnections so that
\begin{equation}
F_k = \lambda \omega^{-2},
\end{equation}
where $\lambda$ is a constant proportional to the mean frequency of reconnections.
For now let us ignore the  dissipation by putting $D_k=0$.
Dissipation of kelvulence is due to either sound emission 
\cite{naz_kelvin}
or due
to a friction with the normal component
\cite{lvov_naz_ladik}. This dissipation acts at very short scales and I will
discuss its role in the end of this paper.
 
\section{Directions of the energy and wavenumber cascades.}

First of all, it is instructive to study directions of the energy
and the waveaction cascades. For this, let us re-write 
equation (\ref{forth}) in two different forms:
a continuity equation for the waveaction,
\begin{equation}
\dot n =-\partial_k \eta = -2 \omega^{1/2} \partial_\omega \eta,
\label{cont_n}
\end{equation}
and a continuity equation for the energy 
\begin{equation}
\omega \dot n =-\partial_k \epsilon = -2 \omega^{1/2} \partial_\omega \epsilon,
\label{cont_en}
\end{equation}
where $\eta$ and $\epsilon$ are the spectral fluxes of the
waveaction and of the energy respectively,
\begin{equation}
\eta =-{C \over 2 \kappa^{10}} \partial_\omega R
\label{eta}
\end{equation}
and
\begin{equation}
\epsilon ={C \over 2 \kappa^{10}} (R- \omega  \partial_\omega R)
\label{epsilon}
\end{equation}
with
\begin{equation}
R=
 n^6 \omega^{21/2} {\partial^2 \over \partial \omega^2} {1 \over n}.
\label{R}
\end{equation}
Note that $\eta=0$ and $\epsilon=$const on the direct cascade
solution (\ref{direct}) and, respectively,
$\eta=$const and $\epsilon=0$ on the inverse cascade
solution (\ref{inverse}). More generally, on
power-law spectra $n_k =k^{\nu}$ we have
$\eta>0$ for $-\infty < \nu < -17/5$ and for
$-1 < \nu < 0$ (and $\eta \le 0$ otherwise),
whereas 
$\epsilon>0$ for $-\infty < \nu < -3$ and for
$-1 < \nu < 0$ (and $\eta \le 0$ otherwise). 
Particularly, if we take spectrum of a sharp reconnection-produced bend,
$n_k =k^{-4}$, then both energy and waveaction cascades are direct,
$\eta>0$ and $\epsilon>0$. This fact is an indication that kelvulence
forced by reconnections should be dominated by the direct cascade
rather than the inverse cascade scalings.
However, the steady state spectrum will be different from
$n_k =k^{-4}$ due to the redistributions of the waveaction and the energy
by the nonlinear wave interactions. Below, we will study such a steady 
state using a reduced version of DAM.

\section{Reduced DAM for Kelvulence forced by reconnections.}

In principle, one can study steady states on kelvulence 
forced by reconnections using DAM as given by equation (\ref{forth}).
However, it is impossible to find a general steady state analytical solution
in this case and one needs to resort to numerics.
On the other hand, most essential details and a full analytical treatment
is possible using a reduced version of DAM,
\begin{equation}
\dot n = {C'  \over \kappa^{10}} \omega^{-1/2} {\partial_\omega}
\left(
 n^4 \omega^{8} {\partial_\omega} (n \omega^{3/2})
\right) + \lambda \omega^{-2}, 
\label{second}
\end{equation}
where $C'$ is an order-one constant.
In this version (for $\lambda=0$) DAM also conserves both the energy and the waveaction
and describes their respective cascade states (\ref{direct}) and (\ref{inverse}), but it 
no longer has thermodynamic Rayleigh-Jeans solutions (\ref{term}).  
Note that the energy and the waveaction  fluxes in this model are respectively
\begin{equation}
\epsilon = - {C'  \over 2 \kappa^{10}} 
 n^4 \omega^{8} {\partial_\omega} (n \omega^{3/2}).
\label{epsilon_2}
\end{equation}
and
\begin{equation}
\eta = - {C'  \over 2 \kappa^{10}} 
 n^4 \omega^{34/5} {\partial_\omega} (n \omega^{17/10}).
\label{eta_2}
\end{equation}
Integrating equation (\ref{second}) once, we get
\begin{equation}
\epsilon=\epsilon_0 - \lambda \omega^{-1/2},
\label{eps_forc}
\end{equation}
where $\epsilon_0$ is a (positive) constant having a meaning of the
asymptotic value of the energy flux at large frequencies.
Integrating one more time we get
\begin{equation}
n = \left( {10 \over C'} \right)^{1/5}   \kappa^{2} 
\omega^{-3/2}
\left(\epsilon_0 \omega^{-1} - {2 \lambda \over 3} 
\omega^{-3/2} - \eta_0 \right)^{1/5},
\label{gen_sol}
\end{equation}
where $\eta_0$ is a (negative) constant having a meaning of the 
asymptotic value of the wavenumber flux at large frequencies.
If there is  no additional (with respect to the reconnections) forcing
then $\eta_0=0$, so that
\begin{equation}
n = \left( {10 \over C'} \right)^{1/5}   \kappa^{2} 
\omega^{-17/10}
\left(\epsilon_0  - {2 \lambda \over 3} 
\omega^{-1/2}  \right)^{1/5}.
\label{gen_sol1}
\end{equation}
At large frequencies, this solution asymptotically approaches to the
direct cascade scaling (\ref{direct}). The inverse cascade scaling
(\ref{inverse}) does not form in any frequency range.
One can relate the asymptotic value of the energy flux
to the minimal frequency of the system $\omega_{min} = \kappa k_{min}^2
= \kappa (2 \pi /L)^2$, where $L$ is the length of the vortex filament.
This relation follows from 
(\ref{eps_forc}) and a the condition that the flux $\epsilon$ is zero
at $\omega_{min} $, i.e.
\begin{equation}
\epsilon_0 = \lambda \omega_{min}^{-1/2}.
\label{min_freq}
\end{equation}
At the minimal frequency, the spectrum tends to a finite value
\begin{equation}
n(\omega_{min}) = \left( {10 \lambda \over 3 C'} \right)^{1/5}   \kappa^{2}
\omega_{min}^{-9/5}.
\label{n0}
\end{equation}
The spectrum (\ref{gen_sol1}) is less steep near $\omega_{min} $
than in the free-cascade range at large $\omega$, but the slope
remains negative for all $\omega$ (i.e. there is no maximum).

\section{Discussion.}

In this paper, I studied the Kelvin wave turbulence (kelvulence) generated by
the vortex reconnections and evolving due to nonlinear wave interactions.
For this, I used the differential approximation model (DAM) of kelvulence
previously introduced in 
\cite{naz_kelvin} and its reduced version (\ref{second}).
The stationary solution of this model (\ref{gen_sol1}) 
describes a state in which the energy flux is directed
toward higher frequencies and it grows from zero at a minimal frequency
$\omega_{min} $ to a constant asymptotic value (\ref{min_freq}) at
large frequencies. In this asymptotic range, the spectrum has
a pure direct cascade scaling
(\ref{direct}). Thus, the answer to the question asked in the beginning of this
paper is  that it is the direct rather than the inverse cascade scaling
that dominates in kelvulence excited by reconnections.
However, a certain amount of waveaction is also produced by the
reconnections per unit time near $\omega_{min} $; it leaks to
smaller frequencies and must be absorbed at the $\omega_{min} $ boundary
(otherwise there would be a pile-up of spectrum near 
$\omega_{min} $ without reaching a steady state).
This absorption seems to arise naturally in the system
because the waveaction conservation takes place only in the weak turbulence
regime which breaks down near $\omega_{min} $, particularly due to the
mode discreteness.

So far we neglected dissipation the role of which is to absorb the
energy cascade at very high frequencies. In superfluids there is no
viscosity and the dissipation is due to either sound generation
by short Kelvin waves (near absolute zero temperature) or due
to a friction with the normal component (at higher temperatures)
\cite{vinen_tur}.
Study of the dissipation effects on the direct cascade in kelvulence
within DAM approach was done in  \cite{naz_kelvin} and  
 \cite{lvov_naz_ladik}. 
It was shown that both of these dissipation mechanisms do not
affect the direct cascade spectrum at low $\omega$ but they arrest
at some large frequency which
results in a sharp cut-off of the spectrum at some maximum frequency
$\omega_{max}$. A finite cut-off due to phonon radiation was 
earlier predicted also in \cite{vinen_sound} and \cite{svist_sound}.

 Assuming that $\omega_{max} \gg \omega_{min}$,
we see that the reconnection forcing and
the radiative/frictional dissipation are separated in 
the frequency space: the forcing is effectively concentrated
near $\omega_{min}$ and the dissipation acts only near
 $\omega_{max}$. This justifies the approach taken in this paper
where we neglected the radiational and frictional radiation
while considering the effect of the reconnection forcing.

An interesting problem for future studies would be 
numerical simulation of DAM given by the fourth-order equation
(\ref{forth}) including the reconnection forcing as well as the
radiation or/and friction dissipation, and comparison of it
with direct numerical simulations of the vortex line
with the same forcing and dissipation mechanism.
It would also be interesting to establish what effects
on kelvulence have curved geometry and non-stationarity of the
vortex filaments.


\begin{thebibliography}{99}

\bibitem{vinen_intro} W.F. Vinen, An introductions to quantum turbulence,
to appear in JLTP (2006).

\bibitem{svist} B.V. Svistunov, Phys. Rev. B {\bf 52}, 3647 (1995).

\bibitem{leb}
V. Lebedev, private communication.

\bibitem{naz_kelvin} S.Nazarenko,  JETP Letters {\bf 83}, 198
(2005).

\bibitem{vinen} W.F. Vinen, Phys. Rev. B {\bf 61}, 1410 (2000).

\bibitem{kivot}
D. Kivotides, J.C. Vassilicos, D.C. Samuels, and C.F.
Barenghi, Phys. Rev. Lett. 86, 3080 (2001). 

\bibitem{ks} E.V. Kozik and B.V. Svistunov, Phys. Rev. Lett. 92,
035301 (2004).

\bibitem{vtm} W.F. Vinen, M. Tsubota and A. Mitani, Phys. Rev. Lett.
91, 135301 (2003).

\bibitem{ks1} E.V. Kozik and B.V. Svistunov, cond-mat/0408241
and Phys. Rev. Lett.,  94, 025301 (2005).


\bibitem{lvov_naz_ladik} V.S. Lvov, S.Nazarenko and L. Skrbek,  Energy
Spectra of Developed Turbulence
 in Helium Superfluids,  arxiv:nlin/0606002 to appear in JLTP (2006).

\bibitem{vinen_tur} W.F. Vinen, Phys. Rev. B {\bf 61}, p 1410 (2000).

\bibitem{vinen_sound} W.F. Vinen, Phys. Rev. B {\bf 64}, 134520 (2001).

\bibitem{svist_sound}  E.V. Kozik and B.V. Svistunov,
 Phys. Rev. B {\bf 72}, 172505 (2005)


\end{thebibliography}
\end{document}